\documentclass[prb,twocolumn,letterpaper,showpacs,preprintnumbers,superscriptaddress,amsmath,amssymb]{revtex4}
\usepackage{graphicx}% Include figure files
\usepackage{dcolumn}% Align table columns on decimal point
\usepackage{bm}% bold math

\begin{document}

\preprint{SAW}

\title{Quantized charge transport through a static quantum dot\\ using a surface acoustic wave}

\author{N. E. Fletcher}
\affiliation{National Physical Laboratory, Queens Road, Teddington TW11 0LW, United Kingdom}
\author{J. Ebbecke}
\affiliation{National Physical Laboratory, Queens Road, Teddington TW11 0LW, United Kingdom}
\affiliation{Cavendish Laboratory, University of Cambridge, Madingley Road, \\Cambridge CB3 0HE, United Kingdom}
\author{T. J. B. M. Janssen}
\affiliation{National Physical Laboratory, Queens Road, Teddington TW11 0LW, United Kingdom}
\author{F. J. Ahlers}
\affiliation{Physikalisch-Technische Bundesanstalt, Bundesallee 100, D-38116 Braunschweig, Germany}
\author{M. Pepper}
\author{H. E. Beere}
\author{D. A. Ritchie}
\affiliation{Cavendish Laboratory, University of Cambridge, Madingley Road, \\Cambridge CB3 0HE, United Kingdom}

\date{\today}

\begin{abstract}
We present a detailed study of the surface acoustic wave  mediated quantized transport of electrons through a split gate device containing an impurity potential defined quantum dot within the split gate channel. A new regime of quantized transport is observed at low RF powers where the surface acoustic wave amplitude is comparable to the quantum dot charging energy. In this regime resonant transport through the single-electron dot state occurs which we interpret as turnstile-like operation in which the traveling wave amplitude modulates the entrance and exit barriers of the quantum dot in a cyclic fashion at GHz frequencies. For high RF powers, where the amplitude of the surface acoustic wave is much larger than the quantum dot energies, the quantized acoustoelectric current transport shows behavior consistent with previously reported results. However, in this regime, the number of quantized current plateaus observed and the plateau widths are determined by the properties of the quantum dot, demonstrating that the microscopic detail of the potential landscape in the split gate channel has a profound influence on the quantized acoustoelectric current transport.
\end{abstract}

\maketitle

\section{Introduction}
Electronic circuits operating by moving single electrons along channels and between quantum dots have been suggested as the basis for new computing technologies \cite{Barnes00}. 
Some of these technologies require local phase coherence (Quantum Computing)\cite{Lo99} and others do not (Cellular Automata)\cite{Snider98} but a common feature of all of them is that their operation necessarily involves moving-electrons interacting with static-electrons. The technologies for producing electrons trapped in both static and moving quantum dots (QDs) are very advanced and both have been able to demonstrate the capture of single electrons. Detailed studies of single QDs have even observed shell structure on the many electron spectrum for dots with a weak tunneling coupling to the surrounding\cite{Tarucha96} and the Kondo effect for dots in the more strongly coupled regime\cite{Goldhaber-Gordon98}. In many cases QDs show similarities with free single atoms and are therefore often referred to as artificial atoms\cite{Kouwenhoven01}.

Quasi zero-dimensional electronic systems have been realized in GaAs/AlGaAs heterostructures in the last decade. One kind of realization has been achieved by wet etching small pillars in a two-dimensional electron gas (2DEG) with subsequent gate metalization \cite{Tarucha96}. With this method QDs have been processed where the number of electrons in the system could be reduced to a few or even to one electron. Also the effect of different shapes of the QDs on the electronic states has been investigated.

Another technique uses metallic Schottky gates to define zero-dimensional systems in a 2DEG \cite{Kouwenhoven97}. These systems have been examined very thoroughly using transport measurements. In particular, by applying phase-shifted AC voltages to the gates, single electron tunneling transport through these static quantum dots -turnstile device- was achieved ten years ago \cite{Kouwenhoven91}. However,  the cycling frequency of these kind of devices is limited roughly to $f=10$~MHz mainly because of the stochastic tunneling time of the electrons and $RC$ times of the electronic circuits. This low frequency of operation severely limits the application of this technology in, for example, metrology where one is concerned with developing a quantum standard of electrical current \cite{Piquemal00}.

A different kind of single electron transport with frequencies of roughly $f=3$~GHz was first reported by Shilton and co-workers\cite{Shilton96a} in 1996. The dynamic potential of a surface acoustic wave (SAW) has been used to transport a constant number of electrons per wave period through a quantum wire formed in a 2DEG. The quantization of the current has been explained as a result of the Coulomb repulsion of the electron in the SAW induced dynamic QDs. The dynamic piezoelectric potential of the traveling SAW induces moving QDs in the entrance of a one dimensional channel. This quantum wire is set to a point beyond conduction pinch-off and therefore a saddle shaped potential hill above the Fermi level of the 2DEG is induced by the negative voltage at the metallic split gate. Due to the electron-electron repulsion in the dynamic quantum dots a fixed number of electrons are transported through the quantum wire resulting in a quantized current $I=n\cdot e\cdot f$. Here $f$ is the frequency of the SAW, $e$ is the elementary electron charge and $n$ is a fixed integer number. 

However the quantization of the acoustoelectric current is not perfect \cite{Talyanskii97,Cunningham99,Janssen00} and the limitation of the precision has been discussed quite controversially \cite{Flensberg99,Maksym00,Robinson01}. The limited precision has been attributed to the lack of thermal equilibration of the transported electrons with the surrounding 2DEG \cite{Fletcher02}. A second factor affecting the quantization is the presence of impurity states in the semiconductor samples which switch between different energy levels. These impurities often plague transport experiments in mesoscopic systems when the switching occurs on a fast time scale and leads to so-called random telegraph signals (RTS). Most importantly, RTS noise often limits precise measurements \cite{Janssen00} and is one of the main difficulties to overcome in order to establish the single electron transport by SAWs as an accurate quantum standard of current.

In this paper we present a detailed study of the interaction of the dynamic potential induced by a SAW with the static potential landscape of a one dimensional channel. The device used in this work was geometrically identical to those used in previous studies with the exception that the channel region of this device contained a quantum dot formed by an impurity potential near the two-dimensional electron gas. The unintentional quantum dot in this device was extremely stable so that detailed measurements could be performed. Apart from this dominating impurity the density of other electrically active impurities was low and therefore RTS noise virtually absent. 

The key result of this work is that we observe a strong interaction between the static quantum dot and the quantized acoustoelectric transport mechanism. We compare our experimental results with the classical model used in Ref.~[\onlinecite{Robinson01}] and also we propose an altered transport mechanism in the regime of moderate SAW power levels.

This paper is organized as follows: After a brief description of the experiment in section \ref{exp_detail} we give a detailed summary of the results in sections \ref{results_first} to \ref{results_last}. In section \ref{discussion} we will discuss possible models which qualitatively explain our results before summarizing the paper in section \ref{summary}.

\section{Experimental Details\label{exp_detail}}
\begin{figure}
	\includegraphics[width=2.5in]{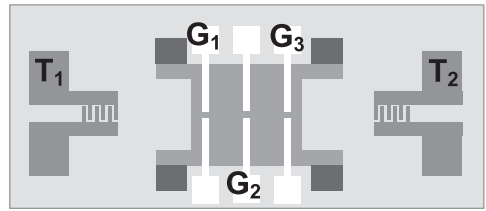}
	\caption{Schematic layout of the sample used in this work. The labeling is explained in the text.}
	\label{fig:sample}
\end{figure}

The device used in this work has been fabricated as described in reference [\onlinecite{Talyanskii98}] where more details of the design are given. The 2DEG is based on a GaAs/ AlGaAs heterojunction and has an electron density of $1.6~\times~10^{11}$~cm$^{-2}$ and a mobility of $1.4\times10^{6}$~cm$^{2}/$Vs (both measured in the dark at $T=1.5$~K). Two SAW interdigital transducers, on either side of a central region containing three split gates, operate at a resonant frequency of 2790~MHz. A schematic sketch of the sample layout is shown in Fig.~\ref{fig:sample}. 

The three geometrically identical metallic split gates have a gap of 800~nm and a length of 800~nm. The dimensions have been chosen to induce channels of roughly 1~$\mu$m length, equal to the wavelength of the SAW. The three split gates labeled G$_{1}$, G$_{2}$ and G$_{3}$ are separated by a distance of 30~$\mu$m and do therefore not influence each other electrically. The two transducers are labeled T$_{1}$ and T$_{2}$. Ohmic source and drain contacts were prepared on both sides of the split gate region with standard alloyed AuGeNi layers and were used to measure the current driven by an applied source-drain voltage or by the surface acoustic wave.

All currents were measured with a Keithley 6514 electrometer. The non-zero burden  voltage of the instrument was compensated by an additional voltage source. Gate voltages were supplied by 18 bit D/A converters which were computer controlled via an optical interface. All measurements were done in DC mode. The sample was mounted in vacuum to the outside wall of a $^{3}$He chamber with a base temperature of 0.3~K. Signals of about 3~GHz were applied to the SAW transducers at generator power levels between -30 and +10~dBm. At levels above approximately 0~dBm the base temperature of the system could not be maintained.

\section{Results}
\subsection{Conductivity of the split gate\label{results_first}}
\begin{figure}
	\includegraphics[width=\columnwidth]{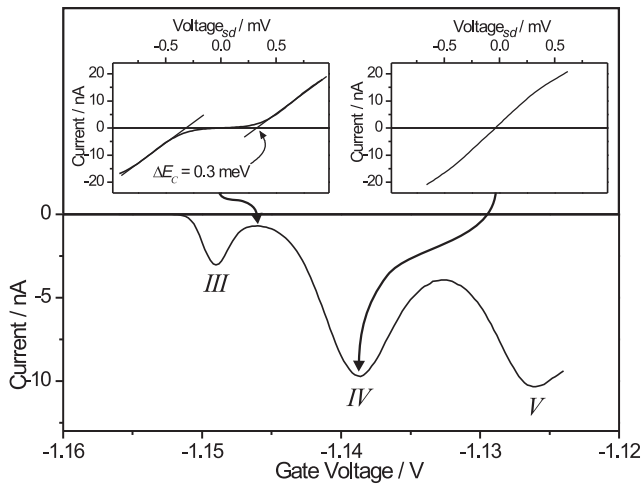}
	\caption{Coulomb blockade oscillations of the electric current close to conduction pinch off of gate G$_{3}$ at source-drain bias 200~$\mu$V, temperature 0.3~K. Upper graphs show the source-drain-voltage dependence of  current measured at the gate voltages indicated by the arrows. Roman numbers labeling the peaks are explained in the text.}
	\label{fig:IV_curves}
\end{figure}

Measurements were performed during three separate cool down cycles of the device, all giving qualitatively similar results. However, the best results were obtained during the second cycle for gate G$_{3}$ and are presented in this paper. The fact that transport properties of a mesoscopic sample vary during different cool downs is a rather common feature\cite{Weis92,Nicholls93}. It illustrates the importance of the microscopic impurity distribution in the sample which is frozen into a different configuration in each cool down cycle. After the second cool down the sample showed clear Coulomb blockade oscillations (CBOs) when the voltage at gate G$_{3}$ was varied in the region of conductivity pinch-off. This is illustrated in Fig.~\ref{fig:IV_curves}. 

The main graph shows the current $I$ through split gate G$_{3}$ as a function of the gate voltage at a constant source-drain voltage $V_{sd}=200 \mu$V. The two upper graphs show $I$-$V_{sd}$ dependencies measured at the points of minimum and maximum conductivity indicated in the figure by arrows. From the typical Coulomb blockade $I$-$V_{sd}$~curve obtained in the conductivity minimum (upper left inset) a Coulomb blockade energy $E_{C}$ of 0.3~meV is estimated. A simple split gate geometry would not give rise to CBOs, therefore we have to conclude that the combination of the random frozen-in impurity potential and the applied split gate potential created a static quantum dot during the second cool down cycle\cite{Nicholls93}. The stability of this dominating impurity (the static quantum dot it induced remained unchanged for more than four weeks) and the virtual absence of other fluctuating background impurities enabled a comprehensive experimental study of the dependence of the acoustoelectric current on RF power and gate voltage.

\subsection{Acoustically driven current\label{results_b}}
\begin{figure}
	\includegraphics[width=\columnwidth]{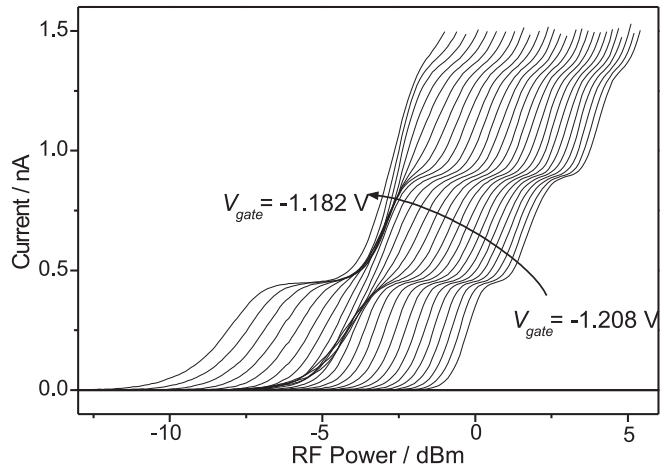}
	\caption{Acoustoelectric current as a function of RF power at different gate voltages. The power refers to the generator output power.}
	\label{fig:fig3}
\end{figure}

With the application of a surface acoustic wave of sufficient power an acoustoelectric current is driven through the split gate even in the region of conductivity pinch-off. A typical set of such measurements is displayed in Fig.~\ref{fig:fig3}.

The measured current is shown as a function of RF power for different settings of the split gate voltage. At high absolute values of gate voltage ($V_{gate}=-1.208$~V) a series of plateaus is observed at current values $I=n\cdot e\cdot f$ for $n = 1,2,3$. On lowering the absolute gate voltage (i.e. on lowering the potential barrier in the split gate) the curves shift to lower RF powers. At some point the $n=1$ plateau vanishes while the $n=2,3$ plateaus are still observed. The $n=1$ plateau 
reappears at lower RF powers but here the $n=2$ plateau is not present and the $n=3$ plateau is very weak. Such behavior has often been seen in quantized acoustoelectric transport experiments in the past, but only occasionally reported\cite{Talyanskii97} without explanation. 
 
\begin{figure}
	\includegraphics[width=\columnwidth]{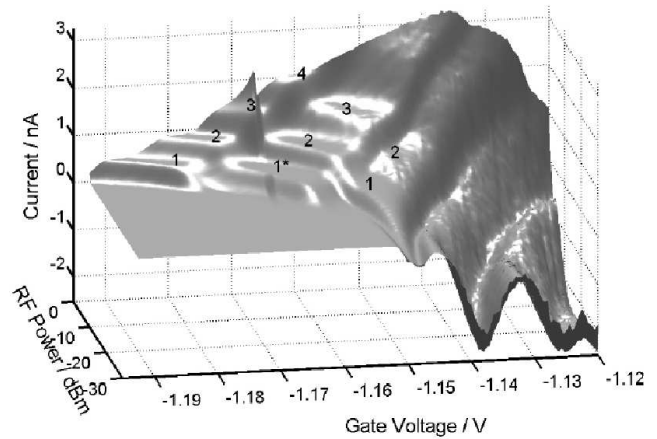}
	\caption{SAW current as a function of gate voltage and RF power presented as a highlighted surface plot. Areas of constant current ('plateaus') are labeled with the number $n$ of electrons transferred per wave cycle. }
	\label{fig:fig3d}
\end{figure}

A complete set of current measurements like those in Fig.~\ref{fig:fig3} was made for RF powers between $-30$~dBm and 0~dBm and for gate voltages from $-1.12$ to $-1.20$~V. At 160 different fixed gate voltages the RF generator power used to excite the SAW transducer T$_{1}$ was swept in steps of 0.2~dB. Sweeping the RF power rather than the gate voltage has the benefit of keeping the average heat load to the sample approximately constant and thus avoids a temperature drift during the measurement time of several hours. The 24000 data points obtained in this way are presented in Fig.~\ref{fig:fig3d} in the form of a three-dimensional surface plot where simulated illumination has been applied to accentuate the structures.

During the whole length of the scan only one switching event disrupted the measurement and it is visible in Fig.~\ref{fig:fig3d} as a discontinuity near $-1.168$~V at high RF power. In the left half of this figure the quantized current plateaus are clearly visible where $n=1, 2$ or 3 electrons (as indicated by the numbers) are transported per SAW cycle. In the right half, the SAW is weaker and the channel is more open, so the current driven by the applied source-drain bias of 50~$\mu$V can be seen. At the lowest applied SAW power of $-30$~dBm, the peaks in the current produced by the source-drain bias occur at the gate voltages corresponding to the conductivity maxima shown in Fig.~\ref{fig:IV_curves}.

The changeover from the acoustoelectric quantized current plateaus to the voltage driven transport regime is not smooth in the sense that the plateaus gradually disappear by getting sloped or 'washed out'.  Instead the current $I$ changes in a terrace-like manner successively decreasing by $1\: e\cdot f$ at two stepwise transition lines. At the first transition line the plateau with $I=1\: e\cdot f$ changes to $I=0$, plateau $2\: e\cdot f$ changes to $1\: e\cdot f$, $3\: e\cdot f$ to $2\: e\cdot f$ and so on, as is indicated in the figure. At a second transition line this pattern is repeated. 

Past observations\cite{Talyanskii97} of "missing" plateaus find a natural explanation when looking at the complete context of the plot of Fig.~\ref{fig:fig3d}. In this figure it is possible to make a scan at fixed gate voltage or RF power such that not all the plateaus in the sequence $n=1,2,3,$.. are present.

It is worth noting that the 'best' (i.e. the flattest) plateau among the plateaus in fig.~\ref{fig:fig3d} does not occur at the highest RF power but it is the one labeled with '1*'. It occurs at an RF power of $-6$~dBm, the lowest value reported in  quantized acoustoelectric transport experiments so far.
 
For gate G$_{3}$ a similar plot has been observed when the opposite transducer T$_{2}$ was used to launch the SAW and similar results have also been obtained during other cool down cycles of this device. However, in all cases the plateau structure was of a poorer quality. On some of the cool downs no quantized acoustoelectric current nor CBOs could be observed in gates G$_{1}$ or G$_{2}$ which from an experimental point of view may suggest a correlation between the occurrence of CBOs and quantized acoustoelectric current.

\subsection{Current derivative plots}
\begin{figure}
	\includegraphics[width=\columnwidth]{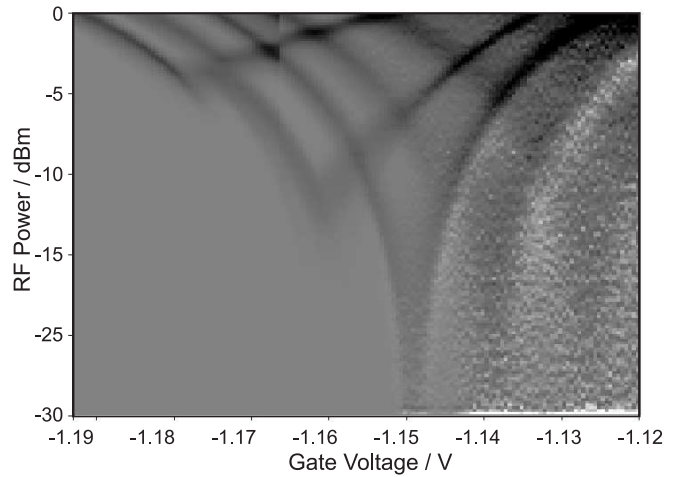}
	\caption{Grey scale coded plot of the numerical derivative $dI/dP_{SAW}$ of current with respect to SAW power based on the data in Fig.~\ref{fig:fig3d}. The medium grey in the lower left corner corresponds to a zero derivative, darker and lighter gray tones indicate positive and negative derivatives.}
	\label{fig:deriv_plot_nonlin}
\end{figure}

The structure in the 3D plot of Fig.~\ref{fig:fig3d} can be better visualized in a two dimensional projection of the data where transitions between plateaus are emphasized by plotting derivatives of the current. This is shown in the grey scale plot in Fig.~\ref{fig:deriv_plot_nonlin}, where $dI/dP_{SAW}$ is displayed. 

\begin{figure}
	\includegraphics[width=\columnwidth]{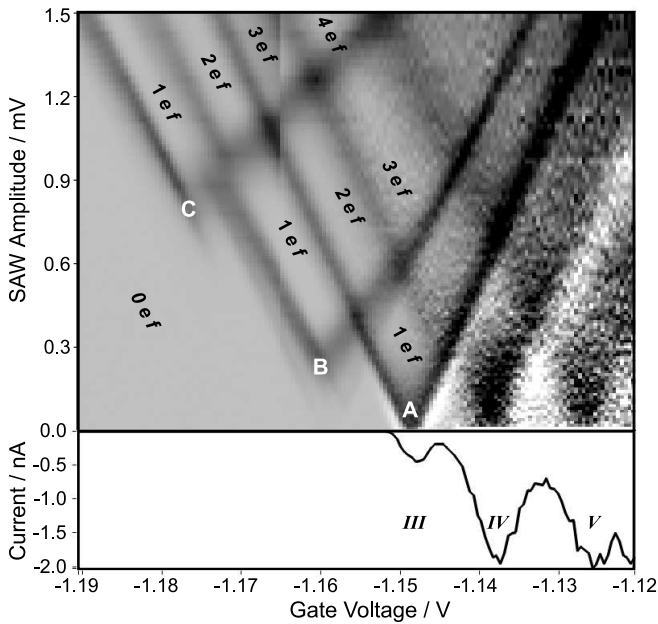}
	\caption{Grey scale coded plot of the numerical derivative $dI/dV_{SAW}$ of current with respect to SAW amplitude based on the data in Fig.~\ref{fig:fig3d}. Light represents small and dark large values of the derivative. The lower graph shows the current at the lowest SAW amplitude.}
	\label{fig:deriv_plot}
\end{figure}

The RF generator used in the experiment only allows changes in the output power in fixed dB increments. However, the SAW potential amplitude is a physically more meaningful quantity and a simpler picture emerges when the change of current with respect to \emph{SAW amplitude} is plotted on a linear voltage scale.
The raw data of fig~\ref{fig:fig3d} and ~\ref{fig:deriv_plot_nonlin} have been linearly interpolated to produce the plot of fig~\ref{fig:deriv_plot}. The amplitude of the SAW induced potential is not easily determined since reflections occurring at the various imperfectly matched transitions in the RF path as well as influences from acoustical reflections occurring on the sample surface and the transducer efficiency are difficult to assess. The amplitude scale of Fig.~\ref{fig:deriv_plot} is therefore derived from an estimate based on the experimental data and on our model which we will outline at the end of section~\ref{model}. The SAW current measured at the lowest RF power is plotted at the bottom of Fig.~\ref{fig:deriv_plot} to demonstrate that the triangular 'fan' patterns visible in the gray-scale derivative plot exactly coincide with the conductivity maxima labeled $III, IV$ and $V$. This strongly suggests that the observed pattern is closely related to the existence of the static QD. 

Figure~\ref{fig:deriv_plot} summarizes the main observation of this paper. It is the highly regular manner in which the quantized acoustoelectric current plateaus develop from the low SAW amplitude, high conductance region in the lower right corner toward the high SAW amplitude and low channel conductance region in the upper left. The structure persists up to the highest SAW amplitudes where the split gate is completely pinched-off. Note that after the last pair of fan lines which originate in a visible conductivity peak (marked by 'A' in Fig.~\ref{fig:deriv_plot}) there are two more fan patterns which start off at the positions marked 'B' and 'C'. 

\subsection{Lateral displacement of the channel}
\begin{figure}
	\includegraphics[width=\columnwidth]{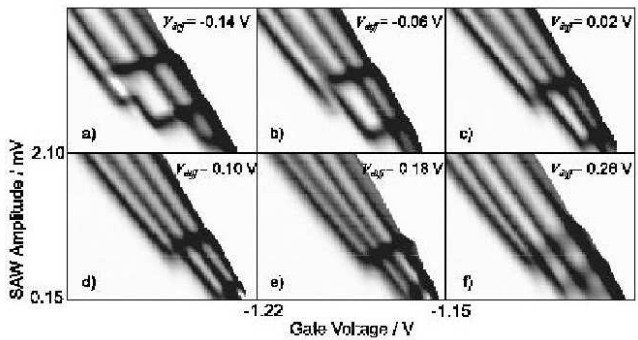}
	\caption{Grey scale coded plot of $dI/dV_{SAW}$ for different voltages applied to the arms of split gate G$_{3}$: $V_{Arm1,2}=V_{G}\pm V_{diff}$, values of $V_{diff}$ are given in the subfigures. SAW amplitude varied from 0.15 to 2.1~mV and gate voltage from -1.15~V to -1.22~V at a temperature of 1.7~K and a frequency of 2787.35~MHz.}
	\label{fig:lateral}
\end{figure}

In the split gate device it is possible to change the interaction between the static quantum dot and the acoustoelectric potential by laterally displacing the induced conductive channel. In the experiment this was achieved by applying, in addition to the main gate voltage, positive and negative voltage offsets to the two arms of the split gate. The main gate voltage was adjusted to maintain the same conductivity of the channel as in the previous measurement. Figure.~\ref{fig:lateral} summarizes a series of measurements for several differential gate voltages. The data are displayed in an identical format to Fig.~\ref{fig:deriv_plot} but in a reduced parameter range.

The series of measurements in Fig.~\ref{fig:lateral} exhibit two main trends: (a) The four plateau separation lines running from upper left to lower right gradually change their spacing and bunch closest together at a differential gate voltage of $\approx+0.10$~V. (b) The two transition lines crossing these separation lines perpendicularly change their position on the vertical amplitude axis, occurring at a lowest amplitude again at approximately $+0.10$~V. The series of measurements displays a symmetric behavior around this differential gate voltage and it is not unreasonable to assume that this is the differential gate voltage where the induced channel is tuned to be aligned with the static quantum dot. Measurements (with no SAW applied) of the periodicity of the CBO peaks as a function of the separate split gate arm voltage~\cite{Weis92,Nicholls93} displayed a similar asymmetry and confirms that the quantum dot is not exactly located in the center of the induced 1D channel.

Two additional observations in Fig.~\ref{fig:lateral} are noteworthy. Firstly, for none of the differential gate voltages does the transition pattern completely disappear.  Secondly, there is an irregular feature occurring at the lowest differential gate voltage of $-0.14$~V where the up to then regular fan structure breaks up and a new current "plateau" emerges, but interestingly with a non-quantized value of $n=0.6$ in the relation $I=n\cdot e\cdot f$. A more detailed measurement of this quasi-plateau is presented in Fig.~\ref{fig:0.6feature}. The lateral displacement measurements strongly support the idea that the observed series of transitions in the acoustoelectric current is related to the static structure existing in the induced 1D channel. At the point where the quantum dot is aligned with the 1D channel, the transitions change rapidly with gate voltage and occur at a low SAW amplitude, suggesting that the interaction is strongest for this geometry.

\subsection{Current dependence on temperature, RF frequency and SAW direction\label{results_last}}
Measurements like those presented in Figs.~\ref{fig:fig3d} to \ref{fig:deriv_plot} have been made at increased temperatures up to a value of 5~K. We observed a thermal smearing of the plateau transition lines in the derivative plots but no other change of the observed fan structure. A quantitative assessment of the temperature effects has not been attempted. A slight variation of the RF frequency around the optimal frequency similarly resulted in no significant change in the fan structure. Upon reversal of the direction of the SAW by using the other IDT we found qualitatively the same pattern which was, however, less well defined.

\section{Discussion\label{discussion}}
\subsection{The classical model}
The quantized acoustoelectric current is thought to result from the dynamic piezoelectric potential of the surface acoustic wave transporting a constant number of electrons per wave period through a one-dimensional channel. For a sufficiently large amplitude the SAW induces moving quantum dots at the entrance of the channel and the quantization of the current has been explained as a result of the Coulomb repulsion of the electrons in the quantum dots. A classical treatment of interacting electrons has been used successfully to explain many qualitative features of the quantized acoustoelectric transport effect\cite{Robinson01}. All models used so far assume a relatively large (of the order of 10~meV) SAW amplitude in a perfectly formed one-dimensional channel. 

The experimental data in Fig.~\ref{fig:deriv_plot} show a pronounced influence of the static QD in the channel on the dynamics of the SAW potential, producing a  systematic structure of transitions between quantized acoustoelectric current plateaus.  In the upper part of the figure at the largest SAW amplitude the current plateaus of the $n=1,2,3,4$ sequence are demarcated by a transition line where, upon lowering the SAW amplitude, the number of transported electrons decreases by one. This could be interpreted as a mechanism by which the static QD captures exactly one electron from the traveling SAW minima per cycle and ejects this electron back into the source reservoir when this line is crossed. We have modeled this behavior using the method in ref.~[\onlinecite{Robinson01}] by including an additional static potential minimum in the synthetic potential landscape of this calculation. By placing the impurity slightly away from the center of the saddle potential maximum toward the driving transducer, transitions between quantized current values could be reproduced by the model at relatively high SAW powers. However, no regular pattern such as shown in Fig.~\ref{fig:deriv_plot} was observed. More importantly, in this configuration the impurity will have no effect on the acoustoelectric current if the SAW direction is reversed by using the opposite transducer. In the experiment we observed a similar fan pattern by using the opposite transducer which can only be explained by having, the unlikely situation, of an identical impurity on the opposite side of the saddle potential. 

The model in ref.~[\onlinecite{Robinson01}] is not easily extended to the low SAW amplitude regime where tunneling through the shallow barriers will become important. Therefore, the observed structure in Fig.~\ref{fig:deriv_plot}, where the transitions lines coincide with the CBO peaks, cannot be modeled by this approach. In this regime an extension of the quantum mechanical approach of ref.[~\onlinecite{Flensberg99}] would be more suitable.

\subsection{The resonant transport model\label{model}}
The key observation that the plateau transition lines in Fig.~\ref{fig:deriv_plot} coincide with the CBO conductivity peaks suggests an alternative explanation which is based on acoustoelectrically driven transfer of electrons through resonant single electron states of the static quantum dot. In many respects it resembles the model of turnstile operation of a semiconductor quantum dot described e.g. in ref. [\onlinecite{Kouwenhoven91}].

\begin{figure*}
		\includegraphics[width=\textwidth]{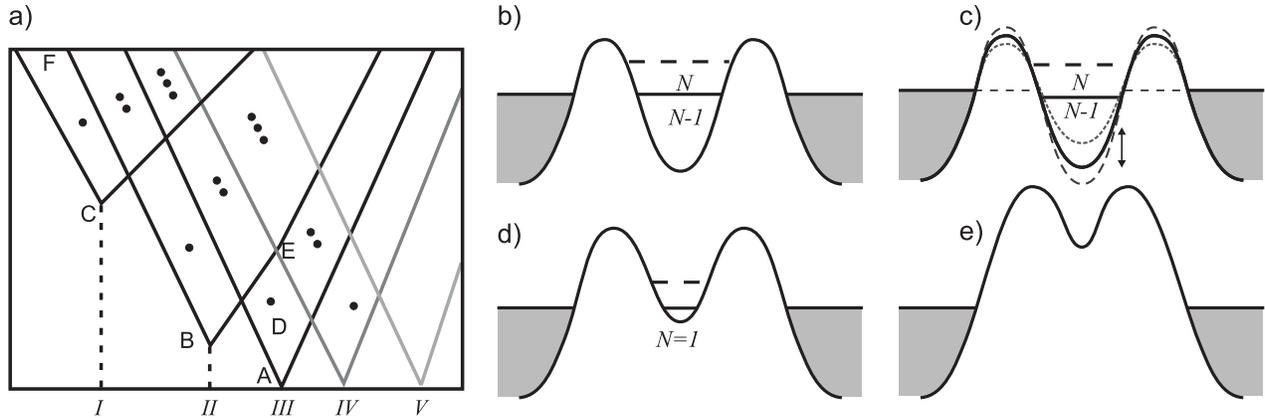}
	\caption{Model of resonant transport of electrons through a static quantum dot. In diagram a) a schematic structure of transition lines as derived from Fig.~\ref{fig:deriv_plot} is shown. In diagrams b) to d) potential profiles along the channel for different total barrier heights are shown. They correspond to the points (A), (D), (C) and (F) in figure a) and are described in the text. Black circles within the rhomboids of diagram (a) indicate the number of electrons transferred per wave period in these regions.}
	\label{fig:model}
\end{figure*}

The resonant transport model is sketched in Fig.~\ref{fig:model} 

where the transition line structure from Fig.~\ref{fig:deriv_plot} is schematically reproduced.
The barrier potential induced by the metallic split gate combined with the potential of a background impurity produces a static quantum dot. The electronic states of this dot when filled with $N$ or $N$-1 electrons, are energetically separated by  an energy $\Delta E=E_{C}+\Delta\epsilon$. Here $E_{C}$ is the  Coulomb charging energy of the quantum dot ($E_{C}=e^{2}/C\approx0.3$~meV, $C$ is the capacity of the quantum dot), and  $\Delta\epsilon$ is the difference between the quantum mechanical eigenstates of the dot for a given number of electrons in the dot (see e.g. ref.~\onlinecite{Kouwenhoven01}). At the temperature of 0.3~K ($kT\approx25~\mu$eV) of our measurements no distinct structure in the CBO peaks was observed and therefore we conclude that $\Delta\epsilon$ is small and does not influence the following qualitative arguments. It will just cause a broadening of the  $E_{C}$-split energy levels, in addition to the thermal broadening. 

In Fig.~\ref{fig:model}(a) some characteristic points are labeled with capital letters. The diagrams (b) to (e) depict the potential landscape along the barrier. Point 'A' at a CBO peak at zero SAW amplitude corresponds to the situation shown in diagram (b), where the single electron state marking the transition from $N-1$ to $N$ electrons in the dot is in resonance with the source and drain Fermi levels at zero source-drain voltage. The conductivity is finite since the barriers are low enough to allow tunneling and a source drain bias or a small amplitude SAW can drive a current. Increasing the SAW amplitude corresponds to moving in diagram (a) from point 'A' to 'D' on a vertical line with the dot state remaining in resonance. For a SAW amplitude large enough to oppose the applied source drain bias, preferentially one electron per SAW cycle is transferred. The SAW modulates the entry and exit barriers to the dot creating a turnstile-like operation. The phase of the entry and exit modulation depends on the barrier separation in relation to the SAW wavelength $\lambda_{SAW}$. If the distance equals $\lambda_{SAW}/2$ the modulation will alternately open the entry and exit barriers. Due to the low potential barriers in this regime, the current is still sensitive to the external bias and therefore the quantization is not perfect. At this SAW amplitude the gate voltage may be changed to move the dot state slightly out of resonance as is indicated in diagram (c). Electron transfer will still occur since during some phase of the SAW cycle the dot state will 'dynamically' come into resonance with the source and drain reservoirs.  

The size of the static dot is not known, but from the measured Coulomb blockade energy of $0.3$~meV we calculate a total capacitance of the dot of $C_{tot}=e^{2}/E_{C}\approx500$~aF. With this value and applying the relation $C=\pi d^2\epsilon\epsilon_{0}/4t$ from ref. [\onlinecite{Smith96}] (where $d$ is the diameter of the dot, $\epsilon_{0}$ the permittivity of free space, $\epsilon$ the relative permittivity of GaAs, and $t=90$~nm the distance of the 2DEG below the surface) a diameter of $0.7$~$\mu$m is estimated. This is slightly larger than half the SAW wavelength of 1~$\mu$m, but still compatible with a turnstile-like action. 

Using this simple model, several features of the observed transition lines in Fig.~\ref{fig:deriv_plot} can be explained. If the SAW amplitude is further increased from point D while keeping the gate voltage constant, when point E is passed, the number of electrons transported per cycle increases from 1 to 3 (as indicated by the numbers of dots within the enclosed plateau regions on the figure). If the gate voltage is set between CBO peaks III and IV, the source and drain fermi levels lie between the electron states in the dot, and the conductance is suppressed. If the SAW amplitude is increased under this condition, the number of electrons transported per cycle increases in the sequence 0,2,4... These sequences of plateaus can be explained if extra electron states within the dot are dynamically involved in the transport as the SAW amplitude is increased. Assuming a symmetric modulation centered around the fermi level, increasing the SAW amplitude will simultaneously access states above and below the fermi level, thereby increasing the number of electrons transported 2 at a time. The alternation of odd and even sequences of plateaus with the variation of gate voltage arises as the lowest available state is aligned or mis-aligned with the fermi level.

There is a striking regularity to the pattern thus produced around CBO peaks $III$, $IV$ and $V$. The diamond shaped regions containing the plateaus are of nearly identical size. In particular, the spacing in SAW amplitude between the crossing points of the lines is constant, and this observation can be used to calibrate the amplitude scale, giving a measure of the size of the SAW potential at the 2DEG. The change in SAW amplitude between points A and E is sufficient to bring states $II$ and $IV$ into the transport in addition to state $III$ and change the current from $I=1\cdot e\cdot f$ to $I=3\cdot e\cdot f$. This amplitude must be approximately equal to the sum of the charging energies $E_{III\rightarrow IV}+E_{II\rightarrow III}$. From the Coulomb Blockade $I$-$V$ curve in the upper left graph of Fig.~\ref{fig:IV_curves} which is for a gate voltage between  between the conductivity peaks $III$ and $IV$ we can deduce $E_{III\rightarrow IV}$ to be approximately $0.3$~meV. The assumption that $E_{III\rightarrow IV}\approx E_{II\rightarrow III}$ is reasonable given the regular size of the diamond regions, so we can conclude $e\cdot A_{SAW}\approx 2E_{III\rightarrow IV}=0.6$~meV at point 'E'. Using this measure, the peak-to-peak amplitude of the SAW induced potential in the measurements of figures \ref{fig:fig3d} to \ref{fig:deriv_plot} ranges from approximately $A_{SAW}\approx 1.5$~mV at 0~dBm generator output to $A_{SAW}\approx 0.05$~mV at $-30$~dBm generator power.

The regular pattern described for CBO peaks $III$, $IV$ and $V$ becomes distorted for gate voltages beyond conductance pinch-off. Two more pairs of transition lines appear at the points labeled 'B' and 'C' in the diagram (a). Although no CBO are visible in the conductance measurements beyond peak $III$, these further transitions lines clearly belong to the sequence so far described, and the inferred CBO peaks $II$ and $I$ have been indicated on the diagram. Even though the barriers have become too high in this region to observe tunneling using a small source-drain bias, with a sufficiently high SAW amplitude it is still possible to use the states within the dot for electron transport. This is the situation illustrated in diagram (d).

The last pair of transition lines is labeled $I$ because no further transitions were observed, even though measurements were extended to RF powers up to $+10$~dBm and correspondingly more negative gate voltages. Beyond point 'C', the plateaus extend in continuous straight regions, to the limit where the heating from the high SAW power prevents further measurements. The plateaus observed in this region with the gate beyond conductance pinch-off show better quantization than for example at point 'D', as the influence of the source-drain bias is reduced, and the current is only determined by the SAW. The point labeled 'F' is in the regime where the original transport model proposed in ref. [\onlinecite{Shilton96}] would apply. Here the impurity QD is raised far above the fermi level of the 2DEG and separated by large barriers, as illustrated in diagram (e). One might expect that the transport at this point would be dominated by the SAW induced moving QDs, but the pattern of transition lines in diagram (a) shows that the number of quantized current plateaus and their widths are still determined by the properties of the static QD.

The accuracy of the quantization of the current transport previously reported for a turnstile device using modulated gates \cite{Kouwenhoven91} was approximately $1~\%$, at an operating frequency of $10~$MHz. The best quantization for the SAW device used here was similar to previous devices, of the order of $0.01\%$, at the much higher operating frequency of $3~$GHz. A simple treatment of the errors of a turnstile considers the tunelling times of wanted and unwanted transport events through the dot in relation to the driving frequency, $f$. These times are determined by the capacitance of the dot, $C$ and the resistances, $R$ of the barriers when raised and lowered during the turnstile cycle. The probability of a tunelling event occuring within an rf cycle is given by $(1-e^{-1/fRC})$.

As the QD in this sample is unintentionally defined in a 1-D channel formed by a single split-gate, it is not possible to control or measure the individual tunnel barrier conductances. The size of the conductance modulation produced by a given SAW power at these barriers is also unknown. However, using the value of $500~$aF estimated above for the capacitance of the dot in this sample, for an error rate of $0.01\%$ at $3~$GHz, the lowered barrier resistance would need to be approximately $70~$k$\Omega$. If the resistance of the barriers without modulation is greater than $1~$G$\Omega$, the unwanted tunnelling throught the dot is suppressed to the $10^{-5}$ level. These estimated values are not inconsistent with turnstile operation of the dot in the region beyond channel pinchoff - in particular the observed accuracy does not require that the lowered barrier conductance is greater than $e^{2}/h$. If this were the case, for part of the rf cycle, the QD would not be defined and isolated from the 2DEG, the energy levels within the dot would no longer be well defined, and the pattern of current quantization would be lost. 

A more complete treatment including non-equilibrium effects within the QD \cite{Liu93} suggests that the error rate should be of the order of $10~\%$ for a turnstile operating at GHz frequencies. It is not clear whether a turnsile model can explain the current quantization over the entire range of barrier heights and SAW amplitudes of the data presented in this paper. If the turnstile model does apply, it is also not clear what the limit on the accuracy of a current standard realised by this approach will be. New experiments will attempt to deliberately define a QD in the path of a SAW using multiple gates. If quantized transport can be replicated using these more controllable QDs, it will be possible to address these questions experimentally.

\subsection{The feature at $0.6\cdot e\cdot f$}
\begin{figure}
	\begin{center}
		\includegraphics[width=3in]{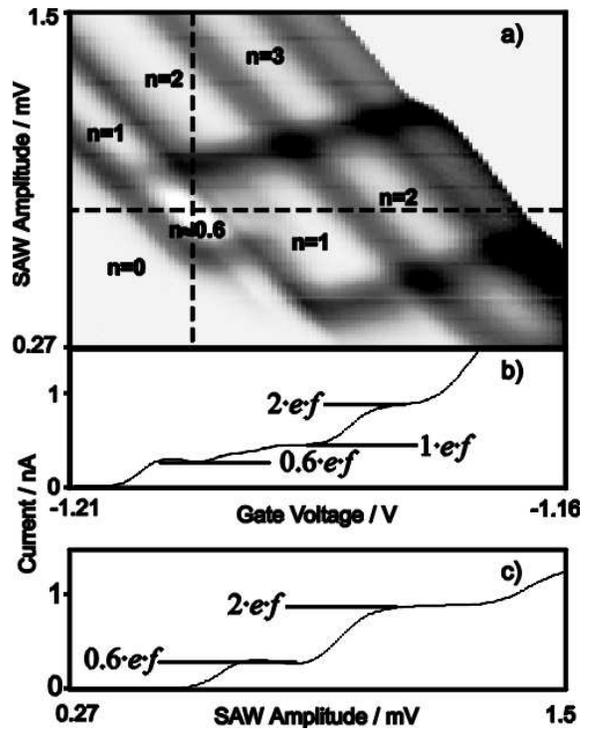}
	\end{center}
	\caption{Grey scale coded plot of $dI/dV_{SAW}$ at a temperature of 0.5~K and a differential voltage of $-0.2$~V (a). Line graphs (b, c) show current scans along the dashed lines to demonstrate the $I=0.6\cdot e\cdot f$ region.}
	\label{fig:0.6feature}
\end{figure}

In Fig.~\ref{fig:0.6feature} one of the frames from Fig.~\ref{fig:lateral} 
is repeated in order to highlight an unexpected feature with a current plateau at a value of $I=0.6\cdot e\cdot f$. 

Additional line scans along gate voltage and SAW amplitude are extracted from the figure and displayed to better demonstrate this plateau. The 0.6-feature forms an area of rectangular shape and is enclosed by the $n=1$ plateaus and the zero-current region. Neither of the simple models presented in this paper predicts this feature or provides an obvious explanation. It exhibits an apparent similarity to the "0.7" structure in ballistic transport measurements of 1D wires, although the physical situation here is quite different. The experimental observation that in 1D wires the "0.7" structure evolves into a "0.6" structure for temperatures between 0.6~K and 3.5~K \cite{Pepper03} does not hold for this plateau. We repeated the measurements as shown in Fig.~\ref{fig:0.6feature} for temperatures between 0.5~K and 10~K. The "0.6" feature was constant at $I=0.6\cdot e\cdot f$ up to 3~K and showed only similar thermal smearing to  the other current plateaus at higher temperatures. At the moment we do not have a satisfactory explanation for the observed 'fractional' plateau. 

\section{Summary and Conclusion\label{summary}}
We have presented detailed measurements of the dependence of the quantized acoustoelectric current on RF power and gate voltage. A regular and reproducible structure of transition lines between multiple quantized current plateaus has been observed for the first time. Each pair of transition lines has its origin in a point which coincides with a Coulomb oscillation peak in the conductivity measurements, which demonstrates a strong influence of the properties of an impurity-defined quantum dot on the electron transport by surface acoustic waves. We propose a new transport mechanism for the quantized acoustoelectric current based on tunneling through a static quantum dot in the low SAW amplitude regime.

The quality of the quantization found in the previous experimental work on quantized acoustoelectric current effects~\cite{Shilton96a,Ebbecke02,Ebbecke00,Talyanskii97,Cunningham99,Cunni00} has not been very reproducible. We have also found that some devices show no quantization of the SAW current at all, even though the split gate induces a channel with clear 1D conductance quantization and the SAW transducers have normal efficiency. The model presented here suggests that the current quantization may rely on an impurity potential inducing a suitable quantum dot, which explains the limited yield of quantized devices. If the SAW current is quantized, the quality of the quantization will depend on the uncontrolled details of the impurity potential, and will be unpredictable.

Further experiments are required to determine whether a static QD is essential to produce a quantized SAW current, or whether the effect can be produced by the dynamic SAW induced QDs in a 'clean' 1D channel, as originally proposed. Future devices with an intentional QD defined in the 2DEG by metallic gates will allow a more systematic study of the interaction of a SAW with a static QD. A thorough understanding of impurity effects and the SAW transport mechanism will be essential if this technology is to be successfully exploited to produce a quantum standard of current or for the much more complex quantum computing applications.

\begin{acknowledgments}
We thank Charles Smith, Varely Talyanskii, Crispin Barnes, Andy Robinson, and James Nicholls for useful discussions. F.J.A wants to thank the Glazebrook Foundation and the Deutsche Forschungsgemainschaft for their support. This work was supported by the National Measurement System Policy Unit of the Department of Trade and Industry, UK
\end{acknowledgments}


\begin{thebibliography}{03}

\bibitem{Barnes00} C. H. W. Barnes, J. M. Shilton, and A. M. Robinson, Phys. Rev. B 62, 8410 (2000)
\bibitem{Lo99} H.-K. Lo, S. Popescu, and T. Spiller, eds., Introduction to Quantum Computation and Information (World Scientific, Singapore, 1999)
\bibitem{Snider98} G. L. Snider, A. O. Orlov, I. Amlani, G. H. Bernstein, C. S. Lent, J. L. Merz, and W. Porod, Semicond. Sci. Technol. 13, A130 (1998)
\bibitem{Tarucha96} S. Tarucha, D. G. Austing, T. Honda, R. J. van der Hage, and L. P. Kouwenhoven, Phys. Rev. Lett. 77, 3613 (1996)
\bibitem{Goldhaber-Gordon98} D. Goldhaber-Gordon, H. Shtrikman, D. Mahalu, D. Abusch-Magder, U. Meirav, and M. A. Kastner, Nature 391, 156 (1998)
\bibitem{Kouwenhoven01} L. P. Kouwenhoven, D. G. Austing, and S. Tarucha, Rep. Prog. Phys. 64, 701 (2001)
\bibitem{Kouwenhoven97} L. P. Kouwenhoven, C. M. Marcus, P. L. MsEuen, S. Tarucha, R. M. Westervelt, and N. S. Windgreen, Mesoscopic Electron Transport 345, 105 (1997)
\bibitem{Kouwenhoven91} L. P. Kouwenhoven, A. T. Johnson, N. C. van der Vaart, and C. J. P. M. Harmans, Phys. Rev. Lett. 67, 1626 (1991)
\bibitem{Piquemal00} F. Piquemal and G. Genev$\grave{\rm e}$s, Metrologia 37, 207 (2000)
\bibitem{Shilton96a} J. M. Shilton, V. I. Talyanskii, M. Pepper, D. A. Ritchie, J. E. F. Frost, C. J. B. Ford, C. G. Smith, and G. A. C. Jones, Journal of Physics - Condensed Matter 8, L531 (1996)
\bibitem{Talyanskii97} V. I. Talyanskii, J. M. Shilton, M. Pepper, C. G. Smith, C. J. B. Ford, E. H. Linfield, D. A. Ritchie, and G. A. C. Jones, Phys. Rev. B 56, 15180 (1997)
\bibitem{Cunningham99} J. Cunningham, V. I. Talyanskii, J. M. Shilton, M. Pepper, M. Y. Simmons, and D. A. Ritchie, Phys. Rev. B 60, 4850 (1999)
\bibitem{Janssen00} T. J. B. M. Janssen and A. Hartland, Physica B 284, 1790 (2000)
\bibitem{Flensberg99} K. Flensberg, Q. Nui, and M. Pustilnik, Phys. Rev. B 60, R16291 (1999)
\bibitem{Maksym00} P. A. Maksym, Phys. Rev. B 61, 4727 (2000)
\bibitem{Robinson01} A. M. Robinson and C. H. W. Barnes, Phys. Rev. B 63, 165418 (2001)
\bibitem{Fletcher02} N. E. F. Fletcher and T. J. B. M. Janssen and A. Hartland, IEE Proc.-Sci. Meas. Technol. 149, 299 (2002)
\bibitem{Talyanskii98} V. I. Talyanskii, J. M. Shilton, J. Cunningham, M. Pepper, C. J. B. Ford, C. G. Smith, E. H. Linfield, D. A. Ritchie, and G. A. C. Jones, Physica B 249-251, 140 (1998)
\bibitem{Weis92} J. Weis, R. J. Haug, K. V. Klitzing, and K. Ploog, Phys. Rev. B 46, 12837 (1992)
\bibitem{Nicholls93} J. T. Nicholls,J. E. F. Frost, M. Pepper, D. A. Ritchie, M. P. Grimshaw, and G. A. C. Jones, Phys. Rev. B 48, 8866 (1993)
\bibitem{Smith96} C. Smith, Rep. Prog. Phys. 59, 235 (1996)
\bibitem{Shilton96} J. M. Shilton, D. R. Mace, V. I. Talyanskii, Yu Galperin, M. Y. Simmons, M. Pepper, and D. A. Ritchie, Journal of Physics - Condensed Matter 8, L337 (1996)
\bibitem{Liu93} C. Liu and Q. Niu, Phys. Rev. B 48, 18320 (1993)
\bibitem{Pepper03} M. Pepper, private communication (2003)
\bibitem{Ebbecke02} J. Ebbecke, K. Pierz, and F.J. Ahlers, Physica E 12, 466 (2002)
\bibitem{Ebbecke00} J. Ebbecke, G. Bastian, M. Bl{\"o}cker, K. Pierz, and F. J. Ahlers, Appl. Phys. Lett. 77, 2601 (2000)
\bibitem{Cunni00} J. Cunningham, V. I. Talyanskii, J. M. Shilton, M. Pepper, A. Kristensen, and P. E. Lindelof, Phys. Rev. B 62, 1564 (2000)

\end{thebibliography}
\end{document}